\documentclass[letter]{aa} % for the letters 
\usepackage{ulem}
\renewcommand{\arraystretch}{1.1}
\usepackage{color}
\usepackage{array}
\usepackage{url}
\usepackage[bottom]{footmisc}
\usepackage{xcolor}
\usepackage{graphicx}
\usepackage{multirow}
\usepackage{caption}
\usepackage{subcaption}
\usepackage{booktabs}
\usepackage{adjustbox}
\usepackage[toc,page]{appendix}
\usepackage{amsmath}
\usepackage{nccmath}
\usepackage{arydshln} % Este paquete permite crear líneas punteadas
\bibpunct{(}{)}{;}{a}{}{,}

\newcommand{\chan}{\textit{Chandra}\xspace}

\newcommand{\xmm}{XMM--\textit{Newton}\xspace}
\newcommand{\nustar}{\textit{NuSTAR}\xspace}

\newcommand{\Ngc}{{NGC 4656 ULX-1}\xspace}
\newcommand{\ngc}{{NGC~4656}\xspace}

\usepackage{txfonts}
\usepackage[]{hyperref}
\hypersetup{unicode=true, colorlinks=true, linkcolor=[rgb]{0.53, 0.15, 0.34}, citecolor=[rgb]{0.06, 0.2, 0.65}, filecolor=[rgb]{1.0, 0.13, 0.32}, urlcolor=[rgb]{0.53, 0.15, 0.34}}

\begin{document} 

\title{A candidate proton cyclotron feature in the ultraluminous X-ray source NGC 4656 ULX-1}

 \subtitle{}

   \author{Nelson Cruz-Sanchez \inst{1}, 
    Enzo A. Saavedra \inst{2,3},
    Federico A. Fogantini \inst{4,5},
    Federico Garc\'{\i}a \inst{1,5},
    Jorge A. Combi \inst{1,5},     \\
    Matteo Bachetti \inst{6}, 
    Matteo Imbrogno \inst{7,8,9},
    Lara Sidoli \inst{10}, 
    Alessio Marino \inst{7,8, 11}   
   }

    \institute{
    Facultad de Ciencias Astron\'omicas y Geof\'isicas, Universidad Nacional de La Plata, B1900FWA La Plata, Argentina \and 
    Instituto de Astrof\'isica de Canarias (IAC), V\'ia Láctea, La Laguna, E-38205, Santa Cruz de Tenerife, Spain \and
    Departamento de Astrof\'isica, Universidad de La Laguna, E-38206, Santa Cruz de Tenerife, Spain \and
    Space Telescope Science Institute, 3700 San Martin Drive, Baltimore, MD 21218, USA \and
    Instituto Argentino de Radioastronom\'ia (CCT La Plata, CONICET; CICPBA; UNLP), C.C.5, (1894) Villa Elisa, Argentina     \and
    INAF, Osservatorio Astronomico di Cagliari, via della Scienza 5 I-09047 Selargius (CA), Italy \and
    Institute of Space Sciences (ICE, CSIC), Campus UAB, Carrer de Can Magrans s/n, E-08193, Barcelona, Spain \and
    Institut d'Estudis Espacials de Catalunya (IEEC), E-08860 Castelldefels (Barcelona), Spain
    \and 
    INAF--Osservatorio Astronomico di Roma, via Frascati 33, I-00078 Monteporzio Catone, Italy
 \and 
    INAF, Istituto di Astrofisica Spaziale e Fisica Cosmica Milano, via A. Corti 12, I-20133 Milano, Italy \and
    INAF/IASF Palermo, via Ugo La Malfa 153, I-90146 - Palermo, Italy 
    }

\abstract{Ultraluminous X-ray sources represent extreme super-Eddington accretion regimes, and a subset is now known to host highly magnetized neutron stars. However, direct observational probes of their surface magnetic fields remain scarce. In this Letter, we report the detection of a narrow X-ray absorption feature at $3.29\pm0.02$~keV in the \textit{XMM--Newton}/EPIC-pn spectrum of \Ngc. The source exhibits a hard ultraluminous state, while our timing analysis reveals a candidate pulsation at $\sim$0.9736~Hz, with a local significance of $5.5\sigma$ and a pulsed fraction of $\sim11\%$. The feature is robust against changes in continuum modeling and data-selection criteria, retaining a statistical significance of $\gtrsim3\sigma$ in Monte Carlo simulations. Interpreting the absorption as a proton cyclotron resonant scattering feature implies a local magnetic field of $B\sim(6$--$7)\times10^{14}$~G in the line-forming region. This value is consistent with strong magnetic fields anchored near the neutron star surface, even if the large-scale dipole is substantially weaker. Although we discuss electron cyclotron features and atomic transitions as possible alternatives, they appear to be less consistent with the observed phenomenology.}


\titlerunning{A 3.3 keV proton absorption line in \Ngc}
\authorrunning{Cruz-Sanchez et al.}
\maketitle

\section{Introduction}

Ultraluminous X-ray sources (ULXs) are off-nuclear objects with apparent luminosities of $L_{\rm X}\sim10^{39}$--$10^{41}$~erg~s$^{-1}$, exceeding the Eddington limit for typical stellar-mass compact objects \citep{FengSoria2011, Kaaret2017, King_2023}. The consensus attributes this extreme output to super-Eddington accretion onto stellar-mass black holes or neutron stars, facilitated by geometric beaming \citep{Poutanen2007,King2009} and strong radiation-driven winds \citep{Pinto2016,Kosec2018,Kosec2021}. The discovery of coherent X-ray pulsations in several ULXs definitively established that a subset of the population harbors neutron stars \citep{Bachetti2014,Fuerst2016,Israel2017,Carpano2018,Walton2018}.

Cyclotron resonant scattering features (CRSFs) are among the few direct probes of these neutron-star magnetic fields \citep{Meszaros1992,CaballeroWilms2012,Staubert2019}. In Galactic X-ray pulsars, electron CRSFs at tens of kilo-electronvolts commonly indicate surface magnetic fields of order $10^{12}$~G. These features often display harmonics and a strong pulse-phase dependence \citep{Truemper1978,Coburn2002,Staubert2019}. In contrast, proton cyclotron features occur at energies lower by a factor of $\sim$1836 than electron cyclotron lines, reflecting the proton-to-electron mass ratio, and for magnetar-strength fields are therefore expected at a few kilo-electronvolts. A narrow proton absorption line at a few kilo-electronvolts would imply a magnetar-strength field ($\sim10^{14}$~G) at the neutron star surface \citep{HoLai2003ApJ,HardingLai2006}. These considerations motivate phase- and state-resolved searches for CRSFs in ULXs, where Comptonized accretion columns and scattering in optically thick winds can significantly shape the observed spectra \citep{BW2007ApJ...654..435B,Middleton2015,Walton2018}. 

Observationally, hints of proton-cyclotron–like absorption have been reported in a few highly magnetized neutron stars. For example, a striking, phase-dependent feature in SGR~0418$+$5729 implies localized magnetar-strength fields, in sharp contrast to the weak global dipole inferred from its spin-down \citep{Tiengo2013}. During several X-ray bursts of SGR~1806$-$20, an absorption feature near $\sim$5~keV was interpreted as a proton CRSF \citep{Ibrahim2002}, and a transient $\sim$8.1~keV feature has been claimed in the anomalous X-Ray pulsar 1RXS~J170849.0$-$400910 \citep{Rea2003}. The isolated neutron star 1E~1207.4$-$5209 exhibits multiple absorption lines at 0.7–2.8~keV, which have been discussed under both cyclotron and atmospheric (atomic) interpretations \citep{Sanwal2002}. However, within the ULX class, a narrow line at $\sim4.5$~keV in M51 ULX-8 has been debated; if interpreted as a proton cyclotron feature, it implies a magnetar-strength local magnetic field near the neutron star surface \citep{Brightman2018}. A subsequent spectral--timing analysis using the covariance spectrum showed that such a strong field cannot correspond to the large-scale dipole, which is constrained to $B_{\rm dip}\lesssim10^{12}$~G, and instead likely probes higher-order multipolar components anchored close to the surface \citep{Middleton2019MNRAS.486....2M}. 

NGC 4656 ULX-1 resides in the low-metallicity, interacting galaxy \ngc \citep{10.1093/mnras/stu1450}, located at a distance of $\sim$5--8~Mpc based on Tully--Fisher measurements \citep{10.1093/mnras/stu1450,10.1093/mnras/sty1934}. In this Letter, we report the discovery of a narrow absorption feature at $\simeq3.3$~keV in the XMM--\textit{Newton}/EPIC-pn spectrum of \Ngc. We quantify its significance using Monte Carlo simulations, and test its robustness against alternative continuum models and data selection criteria. We interpret this line as a proton cyclotron absorption feature (implying a magnetar-strength field for plausible redshifts), and we also discuss possible non-cyclotron explanations. Finally, we outline observations that could distinguish between cyclotron and non-cyclotron scenarios in this ULX.

\section{Observation and data reduction} \label{sec:datareduction}

\Ngc was observed quasi-simultaneously by \xmm (2021 July 10, 26~ks exposure, ObsID 0891020101) and \nustar (2021 July 09, $\sim$168~ks exposure, ObsID 50760001002). Data from the EPIC (pn, MOS2) and FPMA/B detectors were processed using standard reduction pipelines with SAS v21.0.0 and NuSTARDAS v2.1.4, respectively. Source events were screened for background flaring, barycenter-corrected, and grouped for $\chi^2$ spectral fitting. Comprehensive extraction parameters, background thresholds, and instrumental screening criteria are detailed in Appendices~\ref{app:data_xmm} and \ref{app:data_nus}.

\section{Data analysis}

\subsection{Timing analysis}

We began by examining the timing properties of the \xmm\ observation. The EPIC-pn light curve in the 0.3--10\,keV band shows a steady count rate of $\sim 0.5$\,counts\,s$^{-1}$ with no significant variability on the probed timescales. A coherent pulsation search performed with \texttt{HENDRICS} \citep{hendrics} uncovers a candidate signal at $f \approx 0.9736$\,Hz, reaching a local significance of 5.5$\sigma$ and a pulsed fraction of $11\pm1$\,\%. The signal is accompanied by a large apparent frequency derivative ($\dot{f}\sim -2\times10^{-8}$\,Hz\,s$^{-1}$), most naturally interpreted as orbital Doppler modulation. It is strongest below 3\,keV and becomes undetectable above 5\,keV, consistent with its non-detection in the simultaneous \nustar\ data (3--30\,keV). Full details of the timing analysis, including power-density spectra, $Z^2_2$ maps, and Monte Carlo validation, are provided in Appendices~\ref{app:timing_xmm} and \ref{app:timing_nus}.

\begin{figure}
\centering
    \includegraphics[width=\columnwidth]{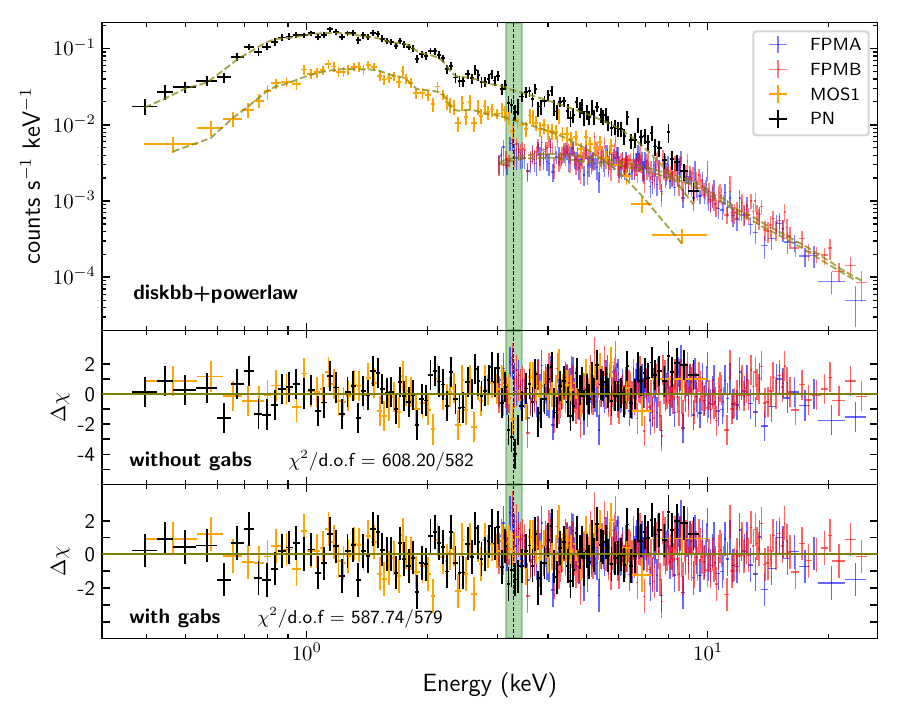}
    \caption{Broadband \xmm\ and \nustar\ spectra of \Ngc. EPIC-pn data are shown in black, MOS2 in orange, FPMA in red, and FPMB in blue. Top panel: Unfolded source spectra over 0.3--25\,keV fit with the {\tt diskbb+powerlaw} continuum. The light green vertical band marks the region around $E_{\rm line}\simeq3.3$\,keV. Middle panel: Residuals of the {\tt diskbb+powerlaw} fit without the absorption component ($\chi^2$/d.o.f. = 608.20/582). Bottom panel: Residuals after including the {\tt gabs} component ($\chi^2$/d.o.f. = 587.74/579). Error bars correspond to 1$\sigma$ uncertainties.}
    \label{fig:fit}
\end{figure}

\renewcommand{\arraystretch}{1.5}
\begin{table}[h]
\centering
\caption{Best-fitting parameters for the {\tt diskbb+powerlaw} and {\tt ThComp(diskbb)} models of the joint pn+MOS2+FPMA/B spectra. $S_{\rm line}$ is the line strength parameter of the {\tt gabs} component (in kilo-electronvolts).}
\label{tab:models_comparison}
\resizebox{\columnwidth}{!}{%
\begin{tabular}{l c c c c}
\toprule
\multirow{2}{*}{\textbf{Parameter}}   & \multicolumn{2}{c}{{\tt Power-law + diskbb}}   & \multicolumn{2}{c}{{\tt ThComp(diskbb)}} \\
\cmidrule(lr){2-3} \cmidrule(lr){4-5}
                                      & {\tt Base} & {\tt + gabs} & {\tt Base} & {\tt + gabs} \\
\midrule
$N_{\rm H,int}$ ($10^{22}$ cm$^{-2}$) & $0.72\pm0.07$ & $0.79^{+0.06}_{-0.07}$ & $0.68^{+0.08}_{-0.09}$ & $0.76\pm0.08$ \\
$\Gamma$ / $\Gamma_{asy}$             & $1.90\pm0.02$ & $1.93\pm0.03$ & $1.91\pm0.02$ & $1.92\pm0.03$ \\
$kT_{e}$ (keV)                        & -- & -- & $80\pm50$ & $80^{+50}_{-50}$ \\
${\rm cov\_frac}$                     & -- & -- & $0.4^{+0.2}_{-0.1}$ & $0.23^{+0.10}_{-0.08}$ \\     
$kT_{seed}$ (keV)                     & $0.15\pm0.01$ & $0.14\pm0.01$ & $0.15\pm0.01$ & $0.14\pm0.01$ \\
Flux ($10^{-12}$ cgs)                 & \multicolumn{2}{c}{$3.48\pm0.05$} & $4.83^{+0.96}_{-0.68}$ & $5.63^{+1.23}_{-0.89}$ \\
\midrule
\textit{Absorption Line}              & & & & \\
$E_{\rm line}$ (keV)                  & -- & $3.29\pm0.02$ & -- & $3.29\pm0.02$ \\
$\sigma$ (keV)                        & -- & $0.04^{+0.03}_{-0.02}$ & -- & $0.04^{+0.03}_{-0.02}$\\
$S_{\rm line}$ (keV)                  & -- & $0.09^{+0.05}_{-0.02}$ & -- & $0.09^{+0.07}_{-0.03}$\\
Line Sig. ($\sigma$)                  & -- & $3.04\pm0.09$ & -- & $2.53\pm0.06$\\
\midrule
$\chi^2$/d.o.f.                       & $608.20/582$ & $587.74/579$ & $605.83/581$ & $586.10/578$ \\
\bottomrule
\end{tabular}%
}
\end{table}

\subsection{Spectral analysis}

Motivated by this candidate pulsation, we performed a detailed spectral analysis of the same dataset with \textsc{XSPEC} \citep{1996ASPC..101...17A}. We fixed the Galactic absorption at $N_{\rm H,Gal}=1.72\times10^{20}\,\mathrm{cm^{-2}}$ using \texttt{tbabs} with the \citet{Wilms2000} abundances, while allowing the intrinsic absorption to vary freely.
We estimated parameter uncertainties with Markov chain Monte Carlo (MCMC) sampling using the \texttt{chain} command in \textsc{XSPEC} \citep[for details, see][]{Fogantini2023,saavedragx13}; all quoted errors are at the 1$\sigma$ level unless stated otherwise.

The 0.3--10\,keV EPIC-pn spectrum is well described by a simple absorbed {\tt powerlaw} model ($\chi^2/\nu = 218.9/192$), though the residuals reveal a clear, narrow absorption feature near $\simeq 3.3$\,keV. Adding a multiplicative Gaussian absorption component ({\tt gabs}) improves the fit by $\Delta\chi^2 = 35.85$ for three additional free parameters. Because standard likelihood-ratio tests (LRTs) can overestimate the significance of spectral-line components, all detection significances were calibrated via $10^5$ Monte Carlo LRT simulations \citep{Protassov2002}. Comparing the null {\tt powerlaw} continuum against the line-included model (Fig.~\ref{fig:sim_lrt}) yields $3.85\pm0.10\,\sigma$ ($p=6\times10^{-5}$) for pn alone. Including the MOS2 spectrum lowers this feature detectability to $3.42\pm0.04\,\sigma$ ($p=3.1\times10^{-4}$); the MOS2 data alone show no significant feature ($0.31\pm0.004\,\sigma$), consistent with their lower effective area near 3.3\,keV.

The minimum-$\chi^2$ solution yields a well-defined centroid and plausible width, yet MCMC sampling reveals a strong width--depth ($\sigma$--$S_{\rm line}$) degeneracy, where broad, shallow lines mimic smooth continuum curvature. Uniform priors were adopted to regularize this non-identifiability: $\sigma \in [0,0.3]~\mathrm{keV}$ (consistent with physical broadening and instrumental resolution) and $S_{\rm line} \in [0,1]~\mathrm{keV}$ (consistent with optically thin lines). These priors yield stable posterior distributions, with line parameters $E_{\rm line}=3.29\pm0.02~\mathrm{keV}$, $\sigma=0.05\pm0.03~\mathrm{keV}$, and $S_{\rm line}=0.12\pm0.03~\mathrm{keV}$. The underlying continuum remains stable: $N_{\rm H,int}=(0.47\pm0.01)\times10^{22}~\mathrm{cm^{-2}}$, $\Gamma=1.77\pm0.03$, and $F_{\mathrm{0.3-10keV}}^{\mathrm{unabs}} = (2.93\pm0.05)\times10^{-12}$~erg~cm$^{-2}$~s$^{-1}$. 

For standard ULX spectral classification, the canonical {\tt diskbb+powerlaw} continuum \citep{Sutton2013MNRAS} was applied to the joint \xmm\ (pn+MOS2) + \nustar\ (FPMA/B) broadband dataset over 0.3--25\,keV (allowing for cross-calibration constants), yielding $\chi^2/\nu=608.20/582$ ($\chi^2_{\nu}\approx1.04$), with $kT_{\rm in}=0.15\pm0.01$\,keV and $\Gamma=1.90\pm0.02$, placing \Ngc\ firmly in the hard ultraluminous (HUL) regime. Adopting a distance of 8.21\,Mpc \citep{10.1093/mnras/sty1934} for consistency with recent optical and spectroscopic studies of the host galaxy, this corresponds to an unabsorbed 0.3--25\,keV luminosity of $L_{\rm X}=(3.20\pm 0.05)\times10^{40}\,{\rm erg\,s^{-1}}$.
The Gaussian absorption component at $\sim$3.3\,keV improves the fit by $\Delta\chi^2=19.49$, corresponding to a detection significance of $3.04\pm0.09\,\sigma$ using Monte Carlo simulations.

A physically motivated thermally Comptonized model, {\tt ThComp(diskbb)}, gives a comparable improvement ($\Delta\chi^2 = 19.73$, $2.53\pm0.06\,\sigma$), with the electron temperature poorly constrained ($kT_{\rm e} = 80\pm50$\,keV) within the available energy range. The line parameters remain stable across all continua tested, confirming that the feature is not an artifact of the continuum choice.

The best-fit parameters for both two-component continuum models (with and without the {\tt gabs} line) are summarized in Table~\ref{tab:models_comparison}. Figure~\ref{fig:fit} shows the unfolded broadband spectra and residuals for the {\tt diskbb+powerlaw} case. To assess whether non-detections in MOS2 and FPMA/B artificially bias the line significance, we reevaluated the joint fit ignoring the 3.1--3.4\,keV bins for these instruments; the EPIC-pn data in this band then yield a revised significance of $3.54\pm0.08\sigma$.

A blind absorption-line scan across the full 0.3--25\,keV range on the joint dataset recovers a single prominent peak at $\sim$3.3\,keV and no other significant features (see Appendix~\ref{app:spectral_nus}). Timing analysis, phase-resolved spectroscopy, and the archival \nustar and \chan data are presented in Appendices~\ref{app:timing_xmm}, \ref{app:nustar}, and \ref{app:chandra}, respectively.

\section{Discussion and conclusions}\label{sec:discussion}

We consistently detect a narrow absorption feature at $\simeq 3.3$~keV in all spectral fits to the simultaneous \xmm\ and \nustar\ observations of \Ngc. Our Monte Carlo likelihood-ratio simulations establish a significance of $3.4$--$3.8\sigma$ for the {\tt power-law} continuum using the EPIC-pn and joint pn+MOS2 spectra. When we add \nustar data up to 25~keV, this significance decreases to $\sim 3\sigma$ for a {\tt diskbb+powerlaw} model, and to $\sim 2.5\sigma$ when we adopt a thermally Comptonized continuum (the {\tt ThComp} model). Although including the full MOS2 and \nustar\ data modestly lowers the significance, excluding the 3.1--3.4 keV range from these instruments restores it to $\sim3.4\sigma$. These results demonstrate that the feature is robust against variations in both the underlying continuum and the instrumental data selection.

The absence of the absorption feature in MOS2 and \nustar\ (FPMA/B) may arise from instrumental limitations: the reduced effective area of MOS2 near 3.3~keV and \nustar's sensitivity above $3$~keV limit the detection of narrow, low-equivalent-width features at $\sim$3.3~keV. Furthermore, while the non-detection in the recent \chan\ observation (see Appendix~\ref{app:chandra}) yields a formal upper limit on the line strength, this constraint remains statistically compatible with the \xmm\ detection. Thus, the \chan\ data cannot unambiguously resolve whether the absence arises from insufficient exposure depth or an intrinsic transient nature of the feature.

The timing analysis provides additional context, though it is still tentative. A coherent pulsation search in the \xmm\ EPIC-pn data identifies a possible candidate signal at $f \approx 0.9736$~Hz that reaches a local significance of $5.5\sigma$ and corresponds to a pulsed fraction of $\sim 11\,\%$. The signal is most prominent below 3 keV and is not detected in the simultaneous \nustar\ observation. Phase-resolved spectra extracted by folding the event list on this candidate period suggest that the absorption feature may be somewhat stronger during the pulse-off phase. However, the current statistics are limited, and deeper observations are needed to confirm this dependence. Given the tentative nature of the timing evidence, we first assessed whether the feature could have an atomic origin (see Appendix~\ref{app:phase_resolved}).

Ultraluminous X-ray sources frequently show highly ionized, fast outflows that produce narrow blueshifted absorption features \citep{Pinto2016,Kosec2018,Kosec2021}. Among atomic transitions in this energy range, the closest match to our measured centroid is S\,\textsc{xvi} at a rest energy of $E\simeq3.276$~keV \citep{vanHoof2018}. Interpreting the feature in this way would require a blueshift of $\sim1300$~km\,s$^{-1}$ to reproduce the observed centroid at $\sim3.29$~keV (see Table~\ref{tab:models_comparison}), a value broadly consistent with wind velocities expected in this class of systems.

However, such an interpretation remains problematic in the absence of additional transitions. A physically consistent photoionized absorber should also imprint companion transitions of comparable or greater strength between 2 and 5~keV, most notably Si\,\textsc{xiv} Ly$\alpha$ (2.006~keV), S\,\textsc{xvi} Ly$\alpha$ (2.62~keV), and Ca\,\textsc{xix} He$\alpha$/Ca\,\textsc{xx} Ly$\alpha$ (3.90--4.11~keV). We do not identify any significant features at these energies. Given the HUL continuum shape inferred for \Ngc, where the funnel is relatively open and wind interception along our line of sight is reduced, the lack of other lines further weakens a purely atomic explanation for \Ngc\ \citep{Middleton2015}, while remaining compatible with the broader ULX wind phenomenology \citep{Pinto2016,Kosec2018,Kosec2021}.

These observations suggest that a proton CRSF offers a more plausible physical interpretation \citep{HoLai2003ApJ,HardingLai2006}. For a fundamental proton cyclotron line, $E_{\rm cyc,p} \simeq 0.63\,\frac{B_{14}}{1+z}\ \mathrm{keV}$,
the observed centroid implies a local magnetic field of $B\simeq(6$--$7)\times10^{14}$~G in the line-forming region for gravitational redshifts $1+z=1.2$--$1.4$, i.e., in the magnetar regime. Strictly speaking, this estimate refers to the near-surface field strength where the feature is produced and does not necessarily coincide with the large-scale magnetic dipole strength.

An important precedent is M51 ULX-8, where a similar narrow feature at $\sim4.5$~keV has been interpreted as a proton CRSF \citep{Brightman2018}. In that source, spectral--timing decomposition of the broadband emission shows that a magnetar-strength dipole is incompatible with the inferred disc geometry and accretion flow, constraining the dipolar field to $B_{\rm dip}\lesssim10^{12}$~G and indicating that the line instead probes a stronger multipolar component close to the neutron star surface \citep{Middleton2019MNRAS.486....2M}. An analogous configuration may apply to \Ngc, where our current data do not yet allow us to distinguish between a strong dipole and a multipole-dominated near-surface field, but do require magnetar-strength fields locally if the 3.3~keV feature is cyclotron in origin.

The modest width ($\sigma\sim0.05$~keV) is consistent with formation close to the surface or in the lower accretion curtain, where thermal and bulk broadening for protons is limited and magnetized radiative transfer can yield relatively narrow, symmetric profiles \citep{LaiHo2002,HoLai2003ApJ}. The absence of an obvious harmonic near $\sim\!6$--7~keV is also expected for proton CRSFs due to suppressed higher-order scattering cross sections \citep{HardingLai2006}. By contrast, an electron CRSF at 3.3~keV would imply $B\approx(2.8$--$3.9)\times10^{11}$~G via $E_{\rm cyc,e}\simeq11.6\,B_{12}/(1+z)$~keV, and electron features typically exhibit broader fundamentals with detectable harmonics at higher energies \citep{Staubert2019}, which we do not observe.

In the HUL geometry, our line of sight likely peers into the accretion funnel, giving an unobstructed view of the accretion column and reduced wind covering \citep{Middleton2015}. Under this scenario, the line centroid appears likely to remain stable (set by the magnetic field strength $B$), while the equivalent width may vary with the accretion rate or geometry (e.g. between HUL and soft ultraluminous states). Higher harmonics would be intrinsically weak or absent in the proton case, as is observed. The candidate pulsed fraction of $\sim11\%$ is also consistent with this picture.
    
In summary, \Ngc\ displays a narrow $\sim$3.3~keV absorption feature whose most natural interpretation is a proton cyclotron line, implying a magnetar-strength local field in the line-forming region. Although an atomic wind absorption line cannot be definitively ruled out, the lack of accompanying transitions and the inferred HUL geometry make such a scenario less compelling in this observation. Together with the candidate proton cyclotron feature in M51 ULX-8 \citep{Brightman2018,Middleton2019MNRAS.486....2M}, our detection in \Ngc\ supports the emerging picture that at least some ULXs host neutron stars with magnetar-level fields in their near-surface layers, while the large-scale dipole may remain in the $\sim10^{11}$--$10^{12}$~G range. Future deep \xmm\ and {\it XRISM} exposures, ideally combined with phase-resolved spectroscopy and hard X-ray coverage, will be crucial to test the persistence of the feature and refine its profile. 

\begin{acknowledgements}
We thank the referee for their constructive comments that improved this manuscript.
We are grateful to the XMM-Newton calibration team for confirming that the feature is not attributable to instrumental or residual calibration inaccuracy.
This project has received funding from the European Union (Project 101183150-OCEANS).
Views and opinions expressed are however those of the author(s) only and do not necessarily reflect those of the European Union or the European Research Executive Agency (REA). Neither the European Union nor REA can be held responsible for them. 
EAS acknowledges support from the Spanish \textit{Agencia estatal de investigaci\'on} via PID2021-124879NB-I00 and PID2024-161863NB-I00. FG and JAC acknowledge support from PIP 0113. JAC was supported by Consejería de Economía, Innovación, Ciencia y Empleo of Junta de Andalucía as research group FQM-322. FG and JAC were also supported by grant PID2022-136828NB-C42 funded by the Spanish MCIN/AEI/ 10.13039/501100011033 and “ERDF A way of making Europe”. M.I. is supported by the ERC Consolidator Grant ``MAGNESIA" (No. 817661) and acknowledges funding from the Catalan grant SGR2021-01269 (PI: Graber/Rea), the Spanish grant ID2023-153099NA-I00 (PI: Coti Zelati), and by the program Unidad de Excelencia Maria de Maeztu CEX2020-001058-M. M.I. is also supported by the ERC Proof of Concept ``DeepSpacePulse" (No. 101189496). MB acknowledges support from Bando Ricerca Fondamentale INAF 2024 Large Grant ``Timing the Ultraluminous Pulsars'' (TULiP).
\end{acknowledgements}

\bibliographystyle{aa}
\bibliography{biblio}

\appendix

\section{\xmm observation}

\subsection{Data reduction}  \label{app:data_xmm}

\begin{figure*}
\centering
\vspace{1.5cm}
    \includegraphics[width=2\columnwidth]{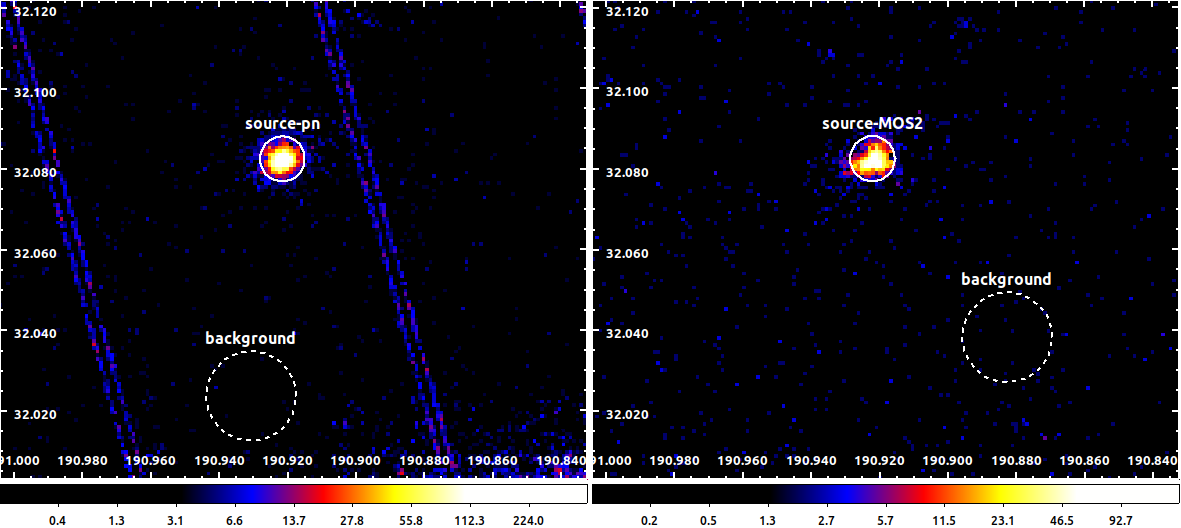}
    \captionsetup{font=small}
    \caption{EPIC-pn (left) and MOS2 (right) 0.3--10 keV images of the field around \Ngc. The source extraction regions (solid circles, 20\arcsec) are centered on the ULX position, while the dashed circles mark the background regions (40\arcsec), chosen on the same CCD and free of contaminating sources. The colorbar scale indicates surface brightness in units of counts per pixel.}
    \label{fig:ds9_display}
\end{figure*}

The \xmm\ observatory observed \Ngc\ on 2021 July 10 (ObsID 0891020101) for a total exposure of 26~ks. The EPIC-pn and MOS cameras were operated in full-frame mode. Data were reduced with SAS v21.0.0, using the tasks {\tt epproc} and {\tt emproc} to produce calibrated event lists. High-background activity intervals were filtered out following standard criteria in the 10--12~keV band, adopting a threshold of 0.35~counts~s$^{-1}$ for MOS and 0.5~counts~s$^{-1}$ for pn, the latter corresponding to the $\sim3\sigma$ upper limit of the $>$10~keV background light curve distribution (100 s time bin) derived from the {\tt lcstats} analysis; no residual flaring was found after filtering. Data from MOS1 were excluded as the source fell on a non-operational CCD. Source events were extracted from a 20\arcsec-radius circle centered on \Ngc\ and background from a nearby source-free 40\arcsec\ region on the same CCD (see Fig.~\ref{fig:ds9_display}). Standard screening was applied ({\tt FLAG==0, PATTERN}$\le4$ for pn; $\le12$ for MOS). Light curves were background-subtracted and exposure-corrected using {\tt epiclccorr}, and barycentered with the {\tt barycen} task. Response matrices and ancillary files were generated with {\tt rmfgen} and {\tt arfgen}. Finally, the spectra were grouped to a minimum 20 counts per bin to allow the use of $\chi^2$ statistics.

\subsection{Timing analysis}\label{app:timing_xmm} 

We searched for variability in \Ngc\ using \xmm\ light curves extracted in the 0.3--10 keV energy band. Visual inspection revealed no significant variability on the probed timescales. The EPIC-pn light curve is consistent with a constant count rate, with an average count rate of $0.5\pm0.1$ counts s$^{-1}$ (1$\sigma$ dispersion).

We investigated the presence of coherent pulsations using the timing analysis tools provided by \texttt{HENDRICS} \citep{hendrics} and \texttt{Stingray} \citep{stingray2019ApJ...881...39H}, analyzing barycenter-corrected EPIC-pn events in the same energy range. As a first step, we explored the aperiodic variability properties of \Ngc\ by computing power density spectra (PDS) in the 0.3--10 keV energy range using different time resolution and segment lengths. In all cases, the overall shape of the PDS is consistent with white (Poisson) noise, with no clear evidence for broad-band noise components or quasi-periodic oscillations.

\begin{figure}[h]
\centering
    \includegraphics[width=\columnwidth]{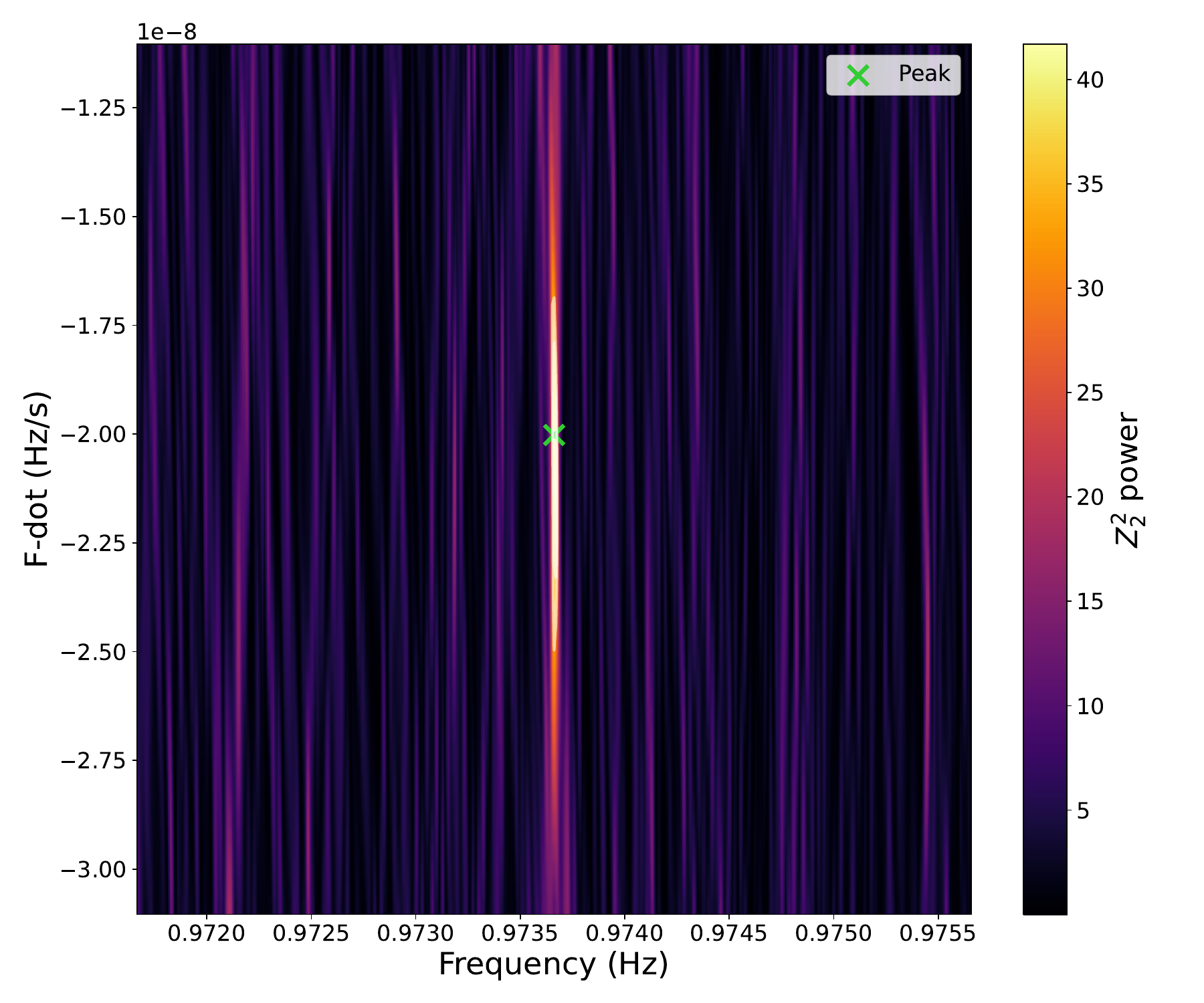}
    \captionsetup{font=small}
    \caption{ Map of the $Z^2_2$ statistic in the frequency ($f$) versus frequency derivative ($\dot{f}$) plane around the candidate signal. The color scale represents the signal power, with the maximum ($Z^2_2 \approx 42$) marked by the green cross.}
    \label{app:freq}
\end{figure}

Although several narrow peaks rise above the average noise level, their statistical significance cannot be established from the PDS alone, and none of them stands out unambiguously as a coherent or quasi-coherent feature across all tested configurations. In particular, a recurrent excess around $\sim$0.6--0.7 Hz is visible in multiple PDS realizations; however, follow-up searches using coherent methods do not confirm this feature as a statistically significant signal. Given the limited sensitivity of PDS-based methods to weak or rapidly evolving coherent signals, we therefore proceeded with a dedicated coherent pulsation search.

We carried out a coherent pulsation search using the acceleration and $Z^2_N$-based techniques implemented in \texttt{HENDRICS}. The analysis was performed on barycenter-corrected EPIC-pn events, exploring trial frequencies in the range 0.01--5 Hz and allowing for a non-zero frequency derivative to account for possible Doppler shifts induced by orbital motion. Several candidate signals were identified in different frequency intervals. Among these, a feature at a frequency of $\sim$0.9736 Hz emerged repeatedly and independently across multiple search configurations, making it the most robust candidate. We subsequently refined the search around this frequency using \texttt{HENzsearch}.

The signal reaches its highest statistical significance for $N=2$, with $Z^2_2 \sim$39--42. Under the null hypothesis of uniformly distributed photon phases, this corresponds to a high local (single-trial) significance of 5.5$\sigma$ ($p=1.94\times 10^{-8}$), exceeding the 99.9\% confidence level based on the theoretical $\chi^2$ distribution of the $Z^2_2$ statistic. We validated this result via Monte Carlo simulations ($10^5$ realizations), finding zero false alarms, which implies a robust empirical lower limit on the significance of $> 4.26\sigma$. We further evaluated the signal using the H-test \citep{de_Jager_2010}, which adaptively compares $Z^2_N$ statistics for different numbers of harmonics and selects the value that maximizes the detection significance. The H-test attains its maximum for $M=2$, indicating that the modulation is best described by two harmonics.

The $Z^2_2$ search landscape in the frequency--frequency derivative ($f$--$\dot{f}$) plane is displayed in Fig.~\ref{app:freq}. The best-fit solution is characterized by a large apparent frequency derivative, $\dot{f}\sim-2\times10^{-8}$ Hz s$^{-1}$. Such a value is unlikely to be intrinsic to the NS spin evolution and is more plausibly interpreted as the result of Doppler modulation induced by orbital motion. Folding the events at the corresponding frequency and frequency derivative yields a modulation consistent with a local pulsed fraction of $11\pm1\%$.

\begin{figure*}[h]
\centering
    \includegraphics[width=\columnwidth]{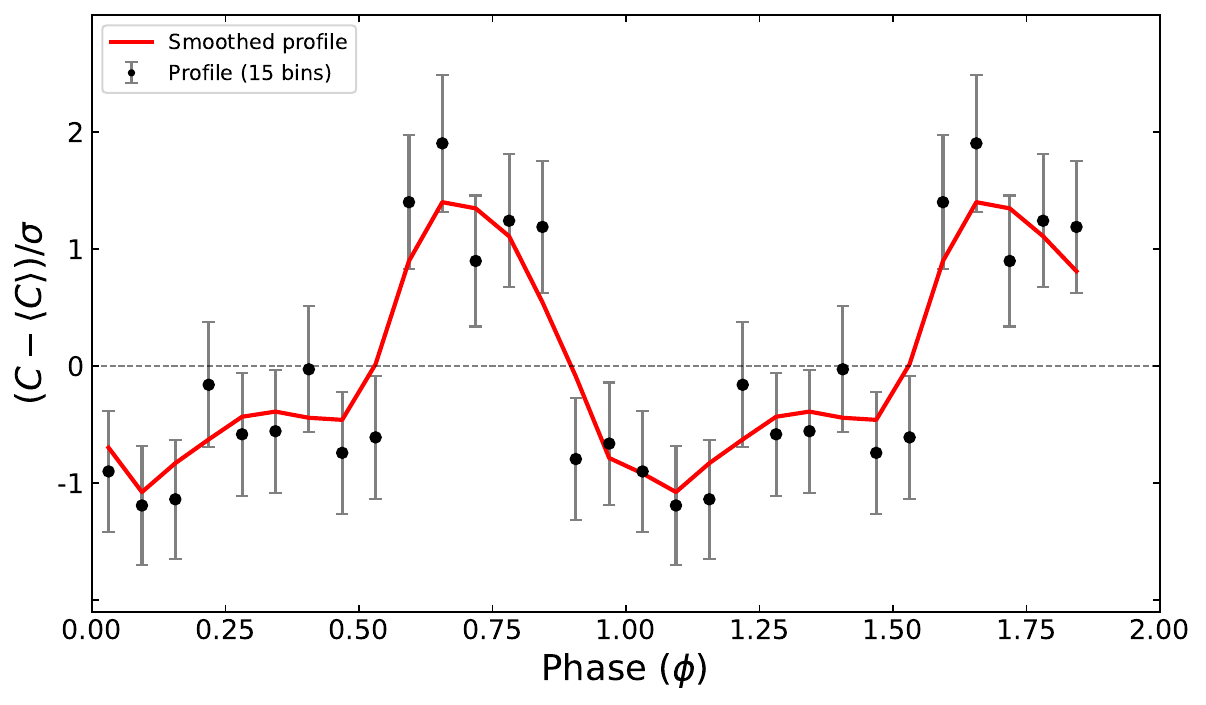}
    \includegraphics[width=\columnwidth]{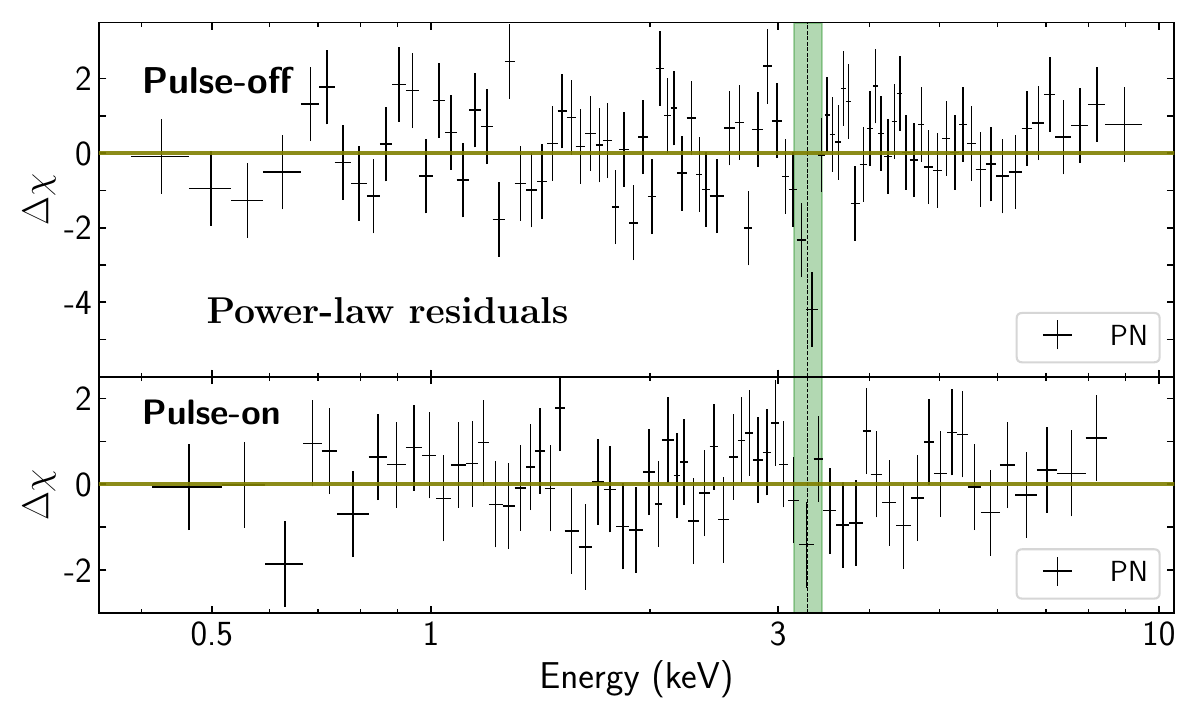}
    \captionsetup{font=small}
    \caption{Left: Folded pulse profile at the candidate pulsation frequency, normalized by subtracting the mean count rate and dividing by the standard deviation. The re-binned profile (15 phase bins) is shown as black points with uncertainties. The red line represents a smoothed version of the folded light curve. The dashed horizontal line indicates the zero-mean level. Right: Residuals ($\Delta\chi$) of the phase-resolved spectra fit with an absorbed power-law model for the pulse-off (top) and pulse-on (bottom) intervals. The shaded green region indicates the expected energy range of the $\sim$3.3 keV absorption feature reported in the phase-averaged analysis.}
    \label{app:fig_pulseprof_spec}
\end{figure*}

We also investigated the energy dependence of the pulsation by splitting the data into three bands. The signal is most significant in the soft band (0.3--3.0 keV: $Z^2_2=31.1$, 4.53$\sigma$). The significance decreases in the medium band (3.0--5.0 keV: $Z^2_2=12.1$, 2.13$\sigma$) and drops below detection thresholds in the hard band (5.0--10.0 keV: $Z^2_2=6.5$, 0.96$\sigma$).

Given the large parameter space explored in the coherent search, the high local significance of the candidate is substantially reduced (to $\approx 2.83\sigma$) when accounting for the effective number of searched trials. We therefore refrain from claiming a firm detection and classify the signal as a promising pulsation candidate requiring independent confirmation with deeper or repeated observations.

An upper limit on the pulsed fraction at the candidate frequency was estimated using the \texttt{HENz2vspf} task, following \citet{Cruz2025a,Cruz2026a}, by comparing the observed $Z^2_2$ statistic with simulated distributions. The simulations were performed assuming a sinusoidal signal with two harmonics ($N=2$), and consisted of 10000 realizations for each pulsed-fraction value. From this procedure, we derive a 90\% confidence upper limit of PF$\lesssim 15\%$ in the 0.3--10 keV energy range.

\subsection{Phase-resolved spectroscopy}\label{app:phase_resolved}

We investigated possible spectral variations of the absorption feature as a function of the candidate pulsation ($\sim$0.9736 Hz, see Appendix~\ref{app:timing_xmm}) by performing a phase-resolved spectral analysis. Using the folded pulse profile obtained with the candidate pulsation, we defined two phase intervals corresponding to relatively lower and higher count-rate portions of the pulse cycle, hereafter referred to as pulse-off ($0\leq\phi<0.55$) and pulse-on ($0.57\leq\phi<0.88$), respectively (see Fig.~\ref{app:fig_pulseprof_spec}).

\begin{table}[h]
\centering
\caption{Phase-resolved spectral fitting results for the pulse-off and pulse-on intervals defined from the candidate pulsation. Fits were performed using an absorbed power-law model ({\tt Base}) and with the addition of a Gaussian absorption line ({\tt + gabs}). Unabsorbed fluxes are reported in the 0.3--10 keV band, with uncertainties quoted at the 1$\sigma$ level.}
\label{app:tab_pulse_fit}
\resizebox{\columnwidth}{!}{%
\begin{tabular}{l c c c c}
\toprule
\multirow{2}{*}{\textbf{Parameter}}   & \multicolumn{2}{c}{{\tt Pulse-off}}   & \multicolumn{2}{c}{{\tt Pulse-on}} \\
\cmidrule(lr){2-3} \cmidrule(lr){4-5}
                                      & {\tt Base} & {\tt + gabs}             & {\tt Base} & {\tt + gabs} \\
\midrule
$N_{\rm H,int}$ ($10^{22}$ cm$^{-2}$) & $0.50\pm0.02$ & $0.50\pm0.02$         & $0.46\pm0.03$ & $0.46\pm0.03$ \\
$\Gamma$                              & $1.88\pm0.04$ & $1.85\pm0.04$         & $1.71\pm0.06$ & $1.70\pm0.06$ \\
Norm ($\times10^{-4}$)                & $4.8\pm0.2$   & $4.8\pm0.2$           & $4.3\pm0.3$   & $4.3\pm0.3$ \\
Flux ($10^{-12}$ cgs)                 & $2.93\pm0.08$ & $2.95\pm0.07$         & $3.0\pm0.1$   & $3.0\pm0.1$ \\
\midrule
\textit{Absorption Line}              & &                                     & & \\
$E_{\rm line}$ (keV)                  & -- & $3.29\pm0.02$                    & -- & $3.26^{+0.08}_{-0.05}$ \\
$\sigma$ (keV)                        & -- & $0.04^{+0.02}_{-0.01}$           & -- & $0.04^{+0.03}_{-0.02}$ \\
$S_{\rm line}$ (keV)       & -- & $0.22^{+0.02}_{-0.07}$           & -- & $0.07^{+0.08}_{-0.05}$ \\
\midrule
$\chi^2$/d.o.f.                       & $114.43/84$ & $89.01/81$              & $47.77/62$ & $45.02/59$ \\
\bottomrule
\end{tabular}%
}
\end{table}

The phase-resolved spectra extracted during the pulse-off and pulse-on intervals are both well described by an absorbed power-law continuum model (Table~\ref{app:tab_pulse_fit}). In the pulse-off phase, structured residuals are observed around $\sim$3.3 keV, and the inclusion of a Gaussian absorption line at this energy provides a statistically improved fit, with line parameters consistent with those measured in the phase-averaged spectrum (see Fig.~\ref{app:fig_pulseprof_spec}). In contrast, during the pulse-on phase the residuals do not show a clear absorption feature in the 3--4 keV band, and the addition of a Gaussian absorption line yields only a marginal improvement, with a line strength consistent with zero within uncertainties. The photon index is slightly harder during pulse-on, while the unabsorbed 0.3--10 keV flux remains consistent between phases and with the phase-averaged value. Overall, these results indicate possible phase-dependent spectral variability, which should be regarded as tentative given the limited statistics of the phase-resolved spectra.

\section{\nustar observation}\label{app:nustar}

\subsection{Data reduction}\label{app:data_nus}

\Ngc was observed simultaneously with \xmm by \nustar on July 09, 2021 (ObsID $50760001002$), for a total exposure of $\sim168$ ks. \nustar consists of two co-aligned focal-plane modules (FPMA and FPMB) operating in the 3--79 keV energy range. Data reduction was performed using NuSTARDAS v2.1.4 within HEASoft v6.34, with calibration files from CALDB v20241126. Calibrated event files were produced with {\tt nupipeline}, applying standard South Atlantic Anomaly filtering ({\tt saacalc=3}, {\tt saamode=strict}, {\tt tentacle=yes}). Source events were extracted from a circular region of 60\arcsec\ radius centered on \Ngc, while background events were selected from a source-free region of the same size. Light curves and spectra were generated using {\tt nuproducts}, and barycentric corrections were applied with {\tt barycorr}. Spectra were grouped to a minimum of 30 counts per bin in the 3--78 keV range to allow the use of $\chi^2$ statistics.

\subsection{Timing analysis}\label{app:timing_nus}

We performed a timing analysis of the \nustar observation of \Ngc using the \texttt{HENDRICS} software package \citep{hendrics}, following procedures consistent with those adopted for the \xmm data (Appendix~\ref{app:timing_xmm}). The aim was to search for the candidate coherent pulsation detected with \xmm at a frequency of $\sim$0.9736 Hz. All event lists were barycenter-corrected and analyzed for FPMA, FPMB, and their combination using \texttt{HENjoinevents}. Unless otherwise stated, the analysis was performed in the 3--30 keV energy range, where \nustar is most sensitive.

We first investigated the aperiodic variability by computing PDS with \texttt{HENfspec}. The PDS, produced for individual modules and combined data, are consistent with pure Poisson noise, with no significant narrow or broad features detected.

Blind coherent pulsation searches were then carried out using \texttt{HENaccelsearch} over the 0.01--5 Hz range, with additional searches extending up to 2 Hz and 10 Hz for completeness. No statistically significant signals were detected in any configuration, and no candidates were found at or near the $\sim$0.9736 Hz frequency identified in the \xmm data.

Motivated by the \xmm results, we performed targeted coherent searches around the candidate frequency using \texttt{HENzsearch}. These searches were conducted for FPMA, FPMB, and combined datasets, exploring different energy bands (3--10 keV, 10--30 keV, and 3--30 keV) and including up to three harmonics ($N=1$--3). No significant signal was detected when using the full $\sim$168 ks exposure.

To account for the possibility of transient pulsations or signal smearing over long integrations, we repeated the targeted searches on shorter data segments of 20, 30, and 60 ks, following an approach similar to that adopted by \citet{Bachetti_2014}. These segment-based analyses also yielded no significant detections. All timing analyses were repeated using different source extraction radii (40\arcsec--60\arcsec), with consistent null results.

Overall, the \nustar timing analysis does not reveal statistically significant coherent pulsations at the candidate frequency of $\sim$0.9736 Hz detected with \xmm, either in blind or targeted searches, for the full exposure or shorter segments, and across different energy bands and extraction regions. We note that this non-detection is fully consistent with the energy dependence observed in the \xmm\ dataset, where the pulsation is significantly detected only at soft energies ($<3$~keV) and becomes statistically insignificant in the hard band ($>3$~keV), effectively placing the signal outside the operational energy range of \nustar.

\begin{figure*}[!ht]
\centering \vspace{2cm}
\begin{subfigure}{0.5\textwidth}
\includegraphics[width=\textwidth]{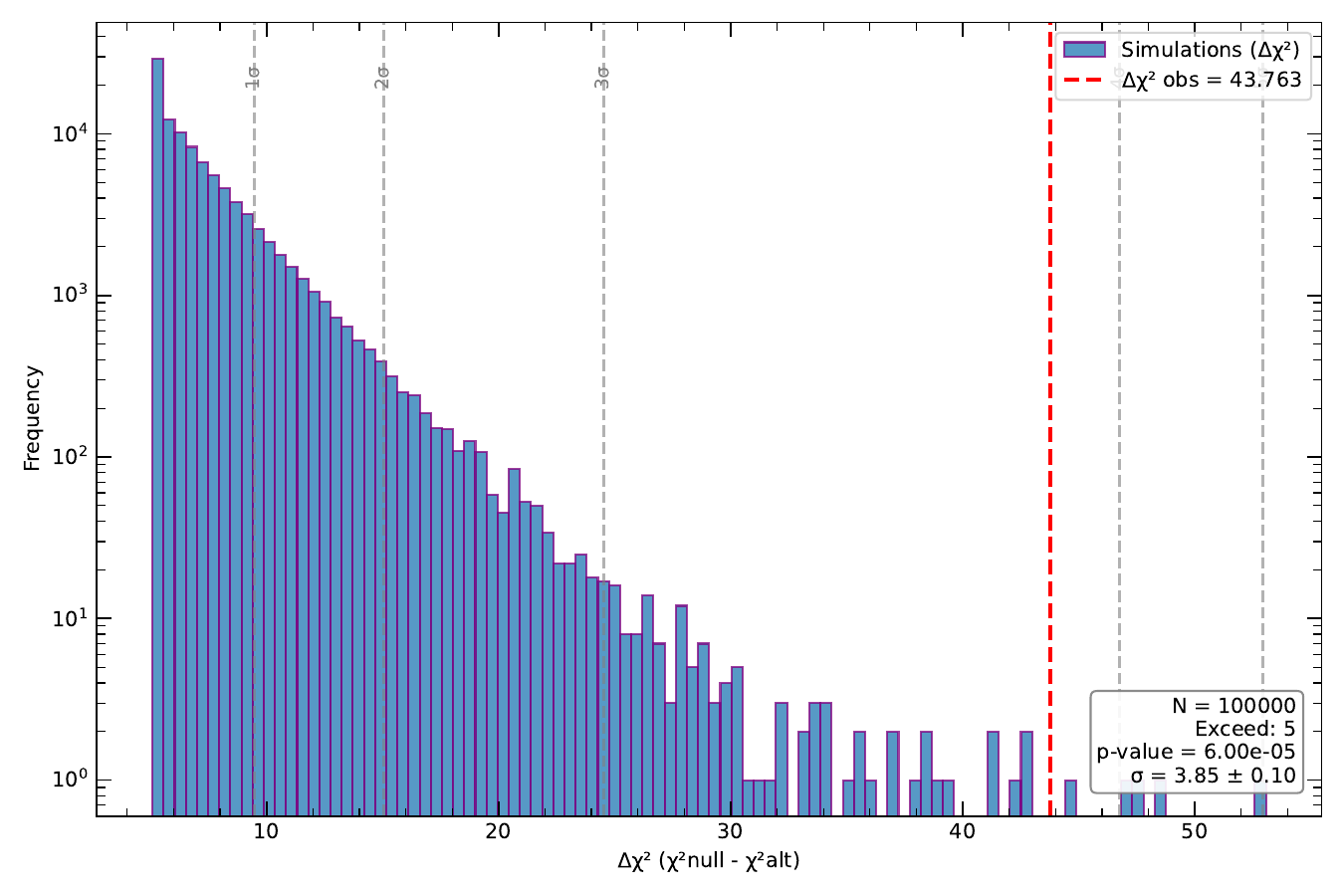}
\caption{{\tt power-law}}
\label{fig:sim_lrt_gabsvspo_pn}
\end{subfigure}\hfill
\begin{subfigure}{0.5\textwidth}
\includegraphics[width=\textwidth]{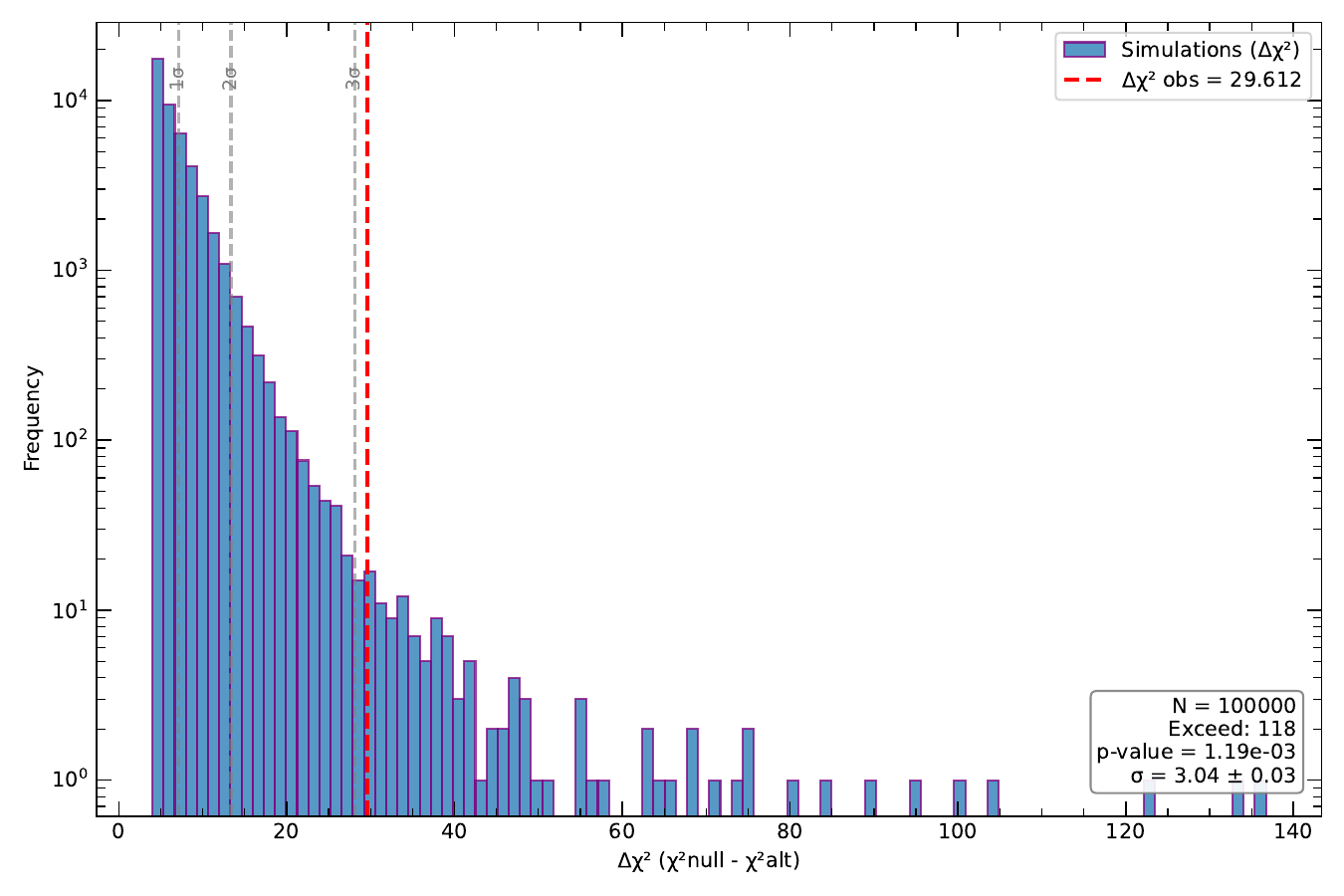}
\caption{{\tt ThComp(diskbb)}}
\label{fig:sim_lrt_gabsvsthcomp_pn}
\end{subfigure}
\caption{\small
Distribution of the simulated $\Delta\chi^2$ values for the EPIC-pn spectra, obtained from $10^5$ Likelihood Ratio Test (LRT) simulations in XSPEC. Left panel: simulations performed under the null model ({\tt powerlaw}), comparing {\tt powerlaw} versus {\tt gabs*powerlaw}. Right panel: simulations performed under the null model ({\tt ThComp(diskbb)}), comparing {\tt ThComp(diskbb)} versus {\tt gabs*ThComp(diskbb)}. In both panels, the red dashed lines mark the observed $\Delta\chi^2$ obtained from the real data when including the multiplicative Gaussian absorption component at $\sim$3.3~keV. The shaded histograms show the simulated $\Delta\chi^2$ distributions under the null hypothesis, while the vertical gray dashed lines indicate the nominal 1--3$\sigma$ thresholds for reference.}
\label{fig:sim_lrt}
\end{figure*}

\subsection{Spectral analysis}\label{app:spectral_nus}

Joint spectral fitting of the \xmm\ (0.3--10 keV) and \nustar\ (3--25 keV) spectra was performed in \texttt{XSPEC}, including multiplicative cross-calibration constants to account for inter-instrument normalization differences, as commonly done in similar \xmm/\nustar\ analyses \citep[e.g.,][]{Saavedra2023A&A...680A..88S}. An absorbed power-law continuum provides an acceptable fit, which improves significantly ($\chi^2/\mathrm{dof}=594.01/581$) with the inclusion of a Gaussian absorption feature at $\sim$3.3 keV. The parameters of this feature are primarily constrained by the \xmm data.

We quantify the impact of the absorption components on the joint \xmm+\nustar\ fit by comparing the {\tt diskbb+powerlaw} continuum against models including one and two multiplicative Gaussian absorption lines. Adding the fundamental line at $\sim$3.3 keV ({\tt gabs*diskbb+powerlaw}) improves the total fit statistic from $\chi^2=608.20$ (582 d.o.f.) to $\chi^2=587.74$ (579 d.o.f.), corresponding to $\Delta\chi^2=20.46$ for three additional free parameters; the corresponding per-instrument contributions are $\Delta\chi^2_{\rm pn}=29.67$, $\Delta\chi^2_{\rm MOS2}=0.36$, $\Delta\chi^2_{\rm FPMA}=-5.21$, and $\Delta\chi^2_{\rm FPMB}=-4.35$. We then searched for a first harmonic by adding a second {\tt gabs} component fixed at twice the fundamental energy (6.58 keV), with its width tied to that of the fundamental and leaving only its strength free to vary. This additional component does not improve the fit statistic: the total $\chi^2$ remains unchanged at 587.74 (579 d.o.f. versus 578 d.o.f.), corresponding to $\Delta\chi^2=0$ for one additional free parameter, with no coherent variation across individual instruments. We therefore derive a 3$\sigma$ upper limit of $0.39$ keV on the harmonic strength.

To assess the significance of the $\sim$3.3 keV feature and to search for additional absorption structures, we performed a blind absorption line scan on the joint \xmm and \nustar spectra. A Gaussian absorption component (\texttt{gabs}) with fixed width of 0.04~keV was added to the best-fitting continuum, stepping the line centroid across the 0.3--25 keV range. At each step, the fit was re-optimized and the corresponding improvement in fit statistic ($\Delta\chi^2$) was recorded.

\begin{figure}[h]
\centering
    \includegraphics[width=\columnwidth]{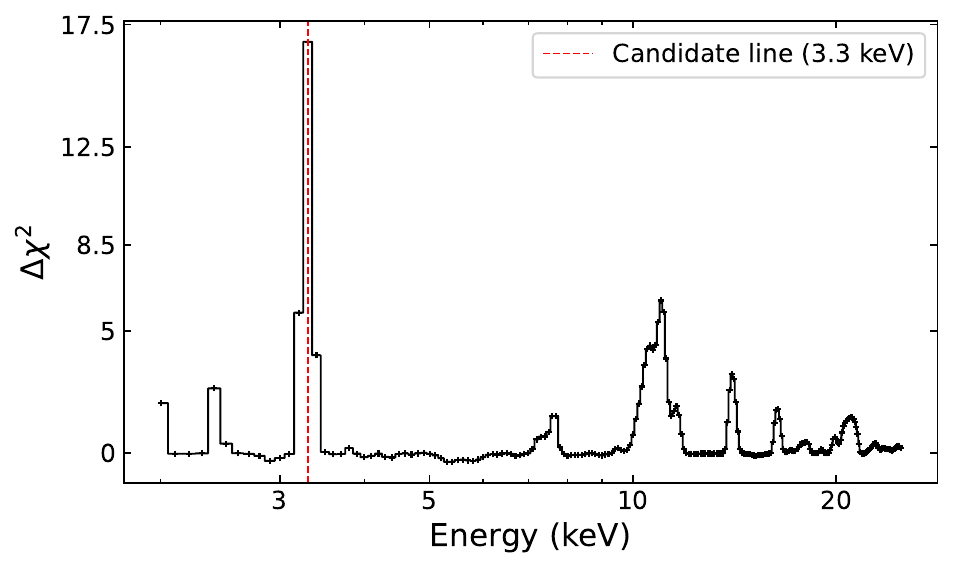}
    \captionsetup{font=small}
    \caption{Blind absorption line scan performed on the joint \xmm (0.3--10 keV) and \nustar (3--25 keV) spectrum. At each trial energy, a Gaussian absorption component with fixed width of 0.04 keV was added to the continuum model, and the improvement in fit statistic ($\Delta\chi^2$) was recorded. A prominent and narrow peak is recovered at $\sim$3.3 keV, corresponding to the candidate cyclotron feature.}
    \label{app:fig_linescan}
\end{figure}

The resulting $\Delta\chi^2$ distribution (Fig.~\ref{app:fig_linescan}) shows a clear and narrow peak at $\sim$3.3 keV, corresponding to the candidate feature identified in the \xmm-only analysis. No comparable excess is detected elsewhere in the scanned energy range, including near the expected position of the first harmonic at $\sim$6.6 keV. A broader excess is present at higher energies around $\sim$11 keV; however, dedicated LRT simulations show that this structure is not statistically significant ($\sim$1.5$\sigma$). Broader fluctuations at higher energies therefore lack the narrow, line-like profile expected for physical absorption features. The blind scan confirms that the $\sim$3.3 keV feature is the most significant line-like structure in the joint spectrum.

\begin{figure*}[!h]
\centering
% --- Fila superior ---
\begin{subfigure}{0.5\textwidth}
\includegraphics[width=\textwidth]{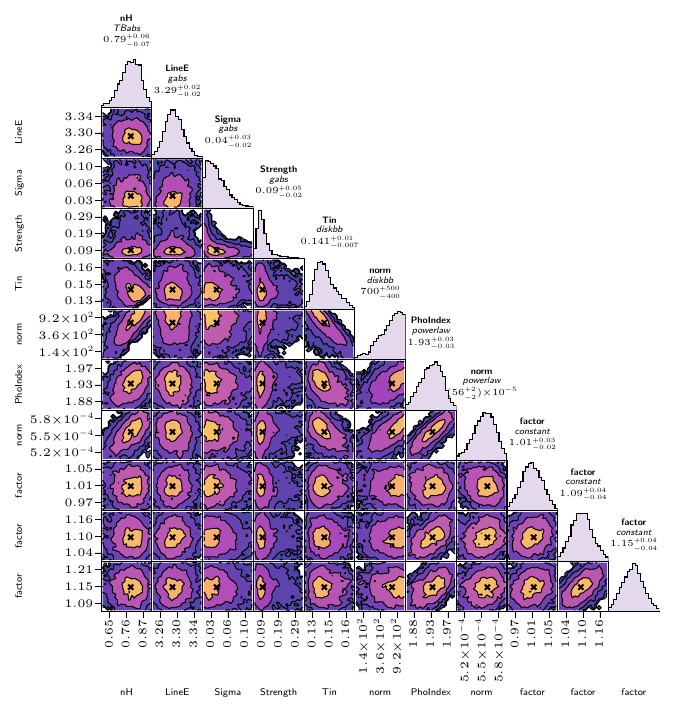}
\caption{{\tt Power-law + diskbb}}
\label{fig:mcmc_pow}
\end{subfigure}\hfill
\begin{subfigure}{0.5\textwidth}
\includegraphics[width=\textwidth]{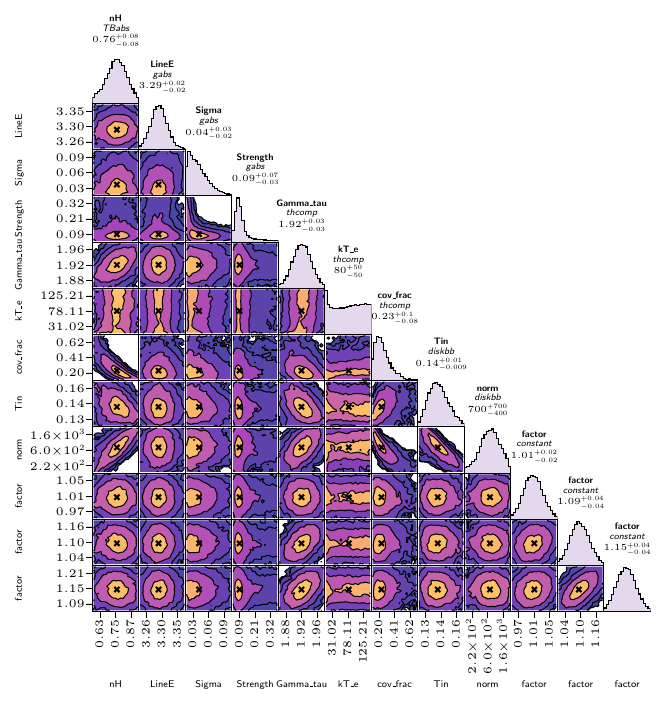}
\caption{{\tt ThComp(diskbb)}}
\label{fig:mcmc_po_diskbb}
\end{subfigure}

\caption{\small
Corner plots showing the posterior probability distributions of the best-fit parameters for the pn+MOS2 detectors obtained from the MCMC analysis of the two tested models with {\tt gabs}: {\tt Power-law + diskbb} (left) and {\tt ThComp(diskbb)} (right). N${\rm H}$ and T${\rm in}$ are expressed in units of $10^{22}$~cm$^{-2}$ and keV, respectively. Each distribution is based on $10^{6}$ converged MCMC iterations.}
\label{fig:mcmc_powgabs}
\end{figure*}

\section{\chan\ observation}\label{app:chandra}

\chan\ observed \ngc\ on April 1, 2025 (ObsID~28114) with a net exposure time of 24.5~ks, using ACIS-I in very faint mode.
The data were reduced using the \chan\ Interactive Analysis of Observation ({\tt CIAO} 4.17) and CALDB 4.12.2, with standard procedures.  
The source spectrum, fit in the 0.5-8 keV energy range, is featureless and well described by an absorbed power-law model ($\chi^2/\nu = 62.74/62$), resulting in the following parameters (1$\sigma$ errors):
$N_{\rm H}=(0.82 ^{+0.28} _{-0.27})\times10^{22}~\mathrm{cm^{-2}}$, 
$\Gamma=1.80^{+0.16} _{-0.15}$, and a flux corrected for the absorption (in the 0.3-10 keV energy range for 
comparison with \xmm\ results), of $F_{\mathrm{0.3-10keV}}^{\mathrm{unabs}}$ = $(1.89 ^{+0.24} _{-0.17})\times10^{-12}$~erg~cm$^{-2}$~s$^{-1}$. 
We then added a narrow Gaussian absorption feature to this continuum model and estimated a 3$\sigma$ upper limit of 0.31 keV on the line strength (fixing the line centroid energy at 3.3 keV and its width at 0.05 keV). The featureless appearance of the \chan spectrum yields a constraint on the line normalization consistent with the absorption line strength measured by \xmm, preventing any firm conclusion about its transient nature.

\section{Supplementary material}\label{app:mcmc}

We present complementary figures supporting the spectral analysis discussed in the main text. Specifically, we show EPIC images illustrating the source and background extraction regions (Fig.~\ref{fig:ds9_display}), the results of Monte Carlo simulations used to assess the statistical significance of the absorption line via likelihood ratio tests (Fig.~\ref{fig:sim_lrt}) and posterior probability distributions from the MCMC exploration of different continuum models including a Gaussian absorption feature (Fig.~\ref{fig:mcmc_powgabs}). These figures provide additional insight into the modeling procedure and illustrate the robustness of the results.

\end{document}